\documentclass[final]{ieee}

\usepackage{graphicx}

\begin{document}
\title{Generation of Test Vectors for Sequential Cell Verification}
\author{Santanu Bhowmick$$ \& Sanjay Bhattacherjee$$ \\ \textit{Indian Statistical Institute} \\ \\ Nandakumar G.N. \\ \textit{ARM Embedded Technologies Pvt. Ltd}.}
\maketitle

\begin{abstract}
For Application Specific Integrated Circuits (ASIC) and System-on-Chip (SOC) designs , Cell - Based Design (CBD) is the most prevalent practice as it guarantees a shorter design cycle ,minimizes errors and is easier to maintain . In modern ASIC design, standard cell methodology is practiced with sizeable libraries of cells, each containing multiple implementations of the same logic functionality , in order to give the designer differing options based on area , speed or power consumtion. For such library cells , thorough verification of functionality and timing is crucial for the overall success of the chip , as even a small error can prove fatal due to the repeated use of the cell in the design. Both formal and simulation based methods are being used in the industry for cell verification. We propose a method using the latter approach that generates an optimised set of test vectors for verification of sequential cells, which are guaranteed to give complete Single Input Change transition coverage with minimal redundancy. Knowledge of the cell functionality by means of the State Table is the only prerequisite of this procedure.
\end{abstract}


\section{Introduction}
The CBD approach using standard cell libraries greatly reduces the dificulties faced while designing custom chips with high performance targets , as compared to transistor level \textit{in situ} customization of cell designs. The Intel Pentium 4 microprocessor is one of the most notable examples of a chip using CBD for a significant portion of its design . But to ensure a successful design , the timing and functionality of the standard cells must be rigorously verified. Though an algorithm for verifying combinational cells is known ~\cite{combin} , currently there is no similar strategy for addressing sequential cells. Here, we pursue a simulation based technique to generate exhaustive test stimuli for any sequential cell given only its State Table (ST). Due to the fact that Single Input Change (SIC) test vectors are more effective than classical Multiple Input Change (MIC) test vectors in terms of delay fault coverage ~\cite{rsic}, we have ensured that the test vector sequence from this method will guarantee 100\% SIC transition coverage.

The paper is organized as follows: Section 2 presents the theoretical concepts that form the basic framework of our approach . Section 3 illustrates the algorithm for test-vector generation. In concluding remarks (Section 4) , possibilities for extending this work is explored.

\section{Preliminary concepts} 
\label{prelim}
\subsection{Single Input Change State Transition Graph}
We define a SIC transition as a single change amongst all input pins and memory elements of the cell. In our approach, the complete functionality of the circuit along with SIC transition information is represented in the form of a modified State Transition Graph , hence forth referred to as the Single Input Change State Transition Graph (SICSTG). This is a directed graph , and every edge in the graph represents a SIC from the source vertex to the destination vertex. The vertices and edges of the SICSTG are defined as follows:-

\subsubsection{Vertex Formation}
\label{vertex_defn}
Any sequential cell can be considered as having $n$ input pins and $m$ output pins, with $k$ internal memory elements(By memory elements , we mean elements that change their output(s) both as a function of the its previous output(s) and the current values of input pins ). We define the \textit{current value} of a input pin or a memory as the digital signal level (0 or 1) on it . By \textit{previous value} , we refer to the logic level on the pin (or memory element) immediately prior to its present logic level . For convenience of reference , the input pins can be further classified into 2 kinds as per its interpretation by the cell - level sensitive and edge sensitive. For the level sensitive pins, the cell considers only the current value of the pin as input to the circuit. But for edge sensitive pins, the immediate previous value at the pin along with the current value is interpreted as an edge input to the circuit . For example, in an edge-triggered D flip-flop, the D input pin is a level sensitive pin but the Clk is an edge sensitive pin.

Any sequential cell has a finite number of defined {\it configurations }, where each configuration is uniquely identified by the following :-
\begin{enumerate}
\item Current values of all level sensitive input pins
\item For each edge - sensitive input pin, its immediate previous value and current value
\item For each internal memory element, its immediate previous value and current value 
\end{enumerate}

We can determine all the configurations of any sequential cell given only its State Table (ST),  each entry in it being a vertex in the SICSTG. The vertex label of each vertex consists of the above 3 tuples in respective order. Take the example of a positive edge-triggered D-type flip-flop:

\begin{table}
\begin{center}
\begin{tabular}{|c|c|c|c|} \hline
$D$ & $Clk$ & $Q_{n}$ & $Q_{n+1}$ \\ \hline
0   &   R   &    0    & 0 \\
0   &   R   &    1    & 0 \\
1   &   R   &    0    & 1 \\
1   &   R   &    1    & 1 \\
0   &   F   &    0    & 0 \\
0   &   F   &    1    & 1 \\
1   &   F   &    0    & 0 \\
1   &   F   &    1    & 1 \\ \hline
\end{tabular}
\caption{State Table for D flip-flop} \label{dffq}
\end{center}
\end{table}

A vertex of SICSTG for this cell would have the following fields in its vertex label: $D$,$Clk_{[n] }$,$Clk_{[n+1]}$,$Q_{[n]}$,$Q_{[n+1]}$ respectively . ( The previous values of any edge-sensitive pin or memory element have the subscript n, whereas the current values have n + 1 ) The edges of the clock are broken down into 2 fields such that for a Rising Edge (R) , \[ Clk_{[n]} = 0 , Clk_{[n+1]} = 1 \] and for a Falling Edge (F), \[ Clk_{[n]} = 1 , Clk_{[n+1]} = 0 \].

For example, the vertex representing the first row of the above ST would have its label as 0, 0, 1, 0, 0 .The rest of the 7 vertex labels could similarly be defined from the ST.

\subsubsection{Edge Formation}
\label{edge_defn}
Two vertices of the SICSTG have a directed edge between them iff the following conditions are satisfied :-
\begin{enumerate}
 \item The previous values of states of the destination vertex must be equal to the respective current values of the source vertex.
 \item The previous values of edge-sensitive input pins of the destination vertex must be equal to the respective current values of the source vertex
 \item There must be only one change in current values of inputs between the source vertex and the destination vertex.
\end{enumerate}

\section{Generation of Test Vectors}
\subsection{Problem Statement}
We need to generate a vector sequence that will verify the functionality as well as timing of any cell given its State Table. This State Table has to be completely specified so that for every possible combination of input \& previous values of state, the values of current states may be ascertained from it. Complete coverage of configurations of the cell for all possibe values of inputs and states of the circuit has to be ensured in order to test the basic functionality of the cell. For timing verification , we need an exhaustive set of test vector that would exercise all possible timing arcs within the circuit at least once. We have used the SIC constraint while generating the test vectors , which is generally believed to provide better delay fault coverage than MIC vectors ~\cite{rsic}. In our algorithm , we generate a set of test vectors along with the expected response of the cell. These input stimuli can be applied to the model under verification and its response compared with that of the generated vectors for complete correctness verification.

\subsubsection{Our Approach}
As per definition , each edge in the SICSTG represents a SIC transition for the given cell. If we can find a walk ~\cite{west} in the SICSTG that uses each edge at least once, then the sequence of vertices in that walk represent a sequence of vectors that cover all SIC transitions in the circuit. In Graph Theory, a walk that covers all edges in a directed graph at least once is known as a Directed Chinese Postman Walk (DCPW) . Once the SICSTG has been created, the problem is transferred to the graph theoretic domain , after which any reasonably efficient implementation of DCPW directly gives us the vectors. We do not discuss the details of the DCPW algorithm as we only use an existing algorithm for determining the ~\cite{CPP}to prove the correctness of our method. The only requirement for the existence of a such a walk is that the SICSTG be strongly connected i.e. there is a path ~\cite{west} from each vertex in the graph to every other vertex in the SICSTG. 

\subsubsection{Vector Generation Algorithm}
The algorithm accepts the State Table as input from the user. The ST must be complete i.e. in case of $n$ inputs and $m$ state elements , the ST must have $2^{m+n}$ entries in it. In case all the inputs are level-sensitive, we can start constructing the SICSTG immediately from the ST. But in case one or more inputs are edge-sensitive ( e.g. the clock input to the circuit) , then we need to expand the ST before proceeding with graph generation. This expansion is necessary to tabulate the behaviour of the circuit for all possible edge-values of the edge-sensitive inputs. For example, the ST supplied might contain entries corresponding to only the Rising Edge (01) and Falling Edge (10) of the clock ( which is an edge-sensitive input pin). But for SICSTG generation, we need to identify the behaviour of the circuit when the clock does not change, i.e. when it recieves 00 or 11.

We have made the following assumption while expanding the ST : 
\textit{If the circuit behaviour is not specified in the ST for a particular combination of input pin values \& previous-values of state elements , then the current-values of state elements remain unchanged on application of this input stimuli }.

This assumption can be justified on the following grounds that were the circuit expected to change its state on application of such an input, the user would have included that entry in the ST. 

The pseudocode for the algorithm is given below.
Let N=Number of level-sensitive inputs , M=Number of edge-sensitive inputs, K=Number of memory elements.

\begin{description}
\item [\textit{\textbf{Step 1:}}] \textit{If $M \geq 1$ i.e. the cell has 1 or more edge-sensitive inputs, then the ST needs to be expanded. We need the ST to be exhaustive i.e. the ST must contain entries for all possible configurations of the cell. In case the behaviour of the cell for a particular combination of values of inputs \& states is not defined in the ST, we assume that for that particular input stimuli, the states of memory elements in the cell remain unchanged , and we insert an entry in the ST corresponding to that. At the end of this step, the ST contains exactly $2^{N+2*M+K}$ entries.}

\item [\textit{\textbf{Step 2:}}] \textit{Each entry in the ST corresponds to a unique configuration of the cell . So, each entry corresponds to a vertex in the SICSTG as described in section \ref{prelim} \ref{vertex_defn}. Thus, there are a total of $2^{N+2*M+K}$ vertices in the SICSTG.}

\item [\textit{\textbf{Step 3:}}] \textit{We form the list of directed edges that constitute the SICSTG as follows - Take a vertex from the list of vertices generated in Step 2 , and find all outgoing SIC edges from it , using the defn of SIC edge from section \ref{prelim} \ref{edge_defn}. Do this for all vertices in the vertex list, to get the complete set of edges of the SICSTG.}

\item [\textit{\textbf{Step 4:}}] \textit{Check if the SICSTG thus generated is strongly connected i.e. check whether all vertices have indegree and outdegree of at least 1. Vertices with indegree or out-degree equal to 0 are deleted from the graph, along with their associated edges. After this step, the SICSTG is guaranteed to be strongly connected.}

\item [\textit{\textbf{Step 5:}}] \textit{To generate an optimal set of SIC test vectors, we need to traverse the SICSTG in such a manner so that all edges are covered at least once. The number of repetitions of edges needs to be minimised to get the minimal set of test vectors. The Directed Chinese Postman Walk on the SICSTG returns a walk containing all the edges at least once, so we use it find such a walk on the SICSTG. This algorithm will always terminate with the optimal vector set as we have ensured that the SICSTG is strongly connected.}
\end{description}

For the D flip-flop given , $D$ is a level-sensitive input, $Clk$ is an edge-sensitive input and $Q$ is a memory element,  so we have $N=M=1$. The complete ST as supplied by the user in Table ~\ref{dffq} is duplicated below, with the edge-sensitive input $Clk$ being broken up into previous-values and current-values.

\begin{table}
\begin{center}
\begin{tabular}{|c|c|c|c|c|} \hline
$D$ & $Clk_{[n]}$ & $Clk_{[n+1]}$ & $Q_{n}$ & $Q_{n+1}$ \\ \hline
0   &   0   &   1   &    0    & 0 \\
0   &   0   &   1   &    1    & 0 \\
1   &   0   &   1   &    0    & 1 \\
1   &   0   &   1   &    1    & 1 \\
0   &   1   &   0   &    0    & 0 \\
0   &   1   &   0   &    1    & 1 \\
1   &   1   &   0   &    0    & 0 \\
1   &   1   &   0   &    1    & 1 \\ \hline
\end{tabular}
\caption{ST of D flip-flop given as input to the algorithm} \label{dffq_2}
\end{center}
\end{table}

After \textit{Step 1} of the algorithm, the entries in Table ~\ref{dffq_new} were added to the ST , as it was found that the ST did not specify the behaviour of the cell on application of these values of inputs as input stimuli.

\begin{table}
\begin{center}
\begin{tabular}{|c|c|c|c|c|} \hline
$D$ & $Clk_{[n]}$ & $Clk_{[n+1]}$ & $Q_{n}$ & $Q_{n+1}$ \\ \hline
1   &   1   &   1   &    0    & 0 \\
0   &   1   &   1   &    0    & 0 \\
0   &   1   &   1   &    1    & 1 \\
1   &   1   &   1   &    1    & 1 \\
1   &   0   &   0   &    0    & 0 \\
1   &   0   &   0   &    1    & 1 \\
0   &   0   &   0   &    0    & 0 \\
0   &   0   &   0   &    1    & 1 \\ \hline
\end{tabular}
\caption{Entries to the ST added in Step 1} \label{dffq_new}
\end{center}
\end{table} 

\begin{figure*}[hbt]
\includegraphics{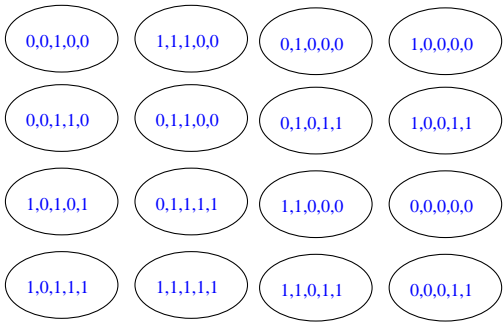} 
\caption{Vertices for D flip-flop}
\label{vertices}
\end{figure*}

Combining Table ~\ref{dffq_2} \& Table ~\ref{dffq_new}, we can now define the list of vertices (Figure ~\ref{vertices}). Each vertex has a vertex label, which stores an unique configuration of the cell . No two vertices can have the same vertex label since the expanded ST has no duplicate entries. The fields in the vertex label are $D$,$Clk_{[n] }$,$Clk_{[n+1]}$,$Q_{[n]}$,$Q_{[n+1]}$ respectively.

\begin{figure*}[hbt]
\includegraphics[width=2.00in]{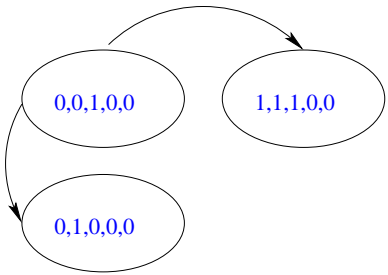} 
\caption{Outgoing edges from vertex labelled $0,0,1,0,0$}
\label{edge}
\end{figure*}

Taking the first vertex from the list, we can list the outgoing SIC edges from it by using the constraints given in Section ~\ref{edge_defn}. The edges satisfying the constraints are listed in Figure ~\ref{edge}. In this way, we can determine all edges in the SICSTG. After the SICSTG for D Flip-flop is completely defined, we check if it is strongly connected. It is found to be so as no vertices have indegree or outdegree as 0. On running the DCPW algorithm ~\cite{CPP} on this graph , we get a walk which covers all edges in the SICSTG at least once. As each edge effectively represents a SIC test vector, we have a sequence of test vectors giving 100\% SIC coverage with minimum repetitions. Vertex-labels of the vertices of SICSTG , written in the order they appear in the walk, form the desired test-vector for the given cell. 

\section{Conclusion}
We have proposed an efficient method to generate an optimal set of SIC test vectors for the verification of sequential cells. The algorithm generates an optimal test vector sequence using a standard implementation of the DCPW algortihm. The time taken by our algorithm is bounded by the time taken to solve the Integer Linear Programming (ILP) problem , which is itself NP-Complete. If a better heuristic is implemented which reduces the worst-case time bound on ILP significantly , then our test vector generation algortihm would also improve by the same degree.
\pagebreak

\end{document}